\documentclass{article}
\usepackage{amssymb}
\usepackage{tikz}
\usetikzlibrary{arrows, calc, shapes, patterns, spy, backgrounds, fit, matrix, positioning}
\usepackage{epstopdf}

\begin{document}

\title{A Calvin bestiary} 

\author{Alan D. Rendall\\
Institut f\"ur Mathematik\\
Johannes Gutenberg-Universit\"at\\
Staudingerweg 9\\
D-55099 Mainz\\
Germany}

\date{}

\maketitle

\begin{abstract}
This paper compares a number of mathematical models for the Calvin cycle of
photosynthesis and presents theorems on the existence and stability of steady 
states of these models. Results on five-variable models in the literature are 
surveyed. Next a number of larger models related to one introduced by 
Pettersson and Ryde-Pettersson are discussed. The mathematical nature of this
model is clarified, showing that it is naturally defined as a system of 
differential-algebraic equations. It is proved that there are choices of 
parameters for which this model admits more than one positive steady state.
This is done by analysing the limit where the storage of sugars from the
cycle as starch is shut down. There is also a discussion of the minimal models
for the cycle due to Hahn.
\end{abstract}

\section{Introduction}\label{intro}

The Calvin cycle is a part of photosynthesis and there are many mathematical 
models for this biochemical system in the literature. Reviews of these can be
found in \cite{arnold11}, \cite{jablonsky11} and \cite{arnold14}. The aim of 
this paper is to survey what is known about the dynamics of these models with 
a focus on what has been proved rigorously. It should be pointed out right 
away that the rigorous results constitute a small island in an ocean of 
simulations and heuristics. To start with it is necessary to fix the 
boundary of the area to be covered. The models treated are all 
deterministic, continuous time evolution equations without delays and spatial 
variations are neglected. Thus mathematically we are dealing with systems of 
ordinary differential equations (ODE) or differential-algebraic equations 
(DAE). The unknowns are concentrations of chemical substances depending on time.

Photosynthesis is a process of central importance in biology and, as a 
consequence, in our daily lives. It consists of two major parts. In the first
of these (the light reactions) energy is captured from sunlight and molecular 
oxygen is produced. In the second (the dark reactions) carbon dioxide from the 
air is used to make carbohydrates. For reasons to be described later the 
second part is also called the Calvin cycle. The models which are the subject
of what follows relate to the Calvin cycle and all describe ordinary 
chemical reactions in solution together with simple sources, sinks and 
transport processes between cellular compartments which fit into the same 
mathematical framework. The light reactions involve electrochemistry on a 
membrane, a type of process whose modelling will not be considered here. A
comprehensive introduction to the biochemistry of photosynthesis can be found
in \cite{heldt11}. 

If we are describing one biological system here, why should there be many
mathematical models for it? This is a consequence of some general features of 
the modelling of biochemical systems which will now be listed. The first is 
that a biochemical system like the Calvin cycle is in reality coupled to many 
other chemical processes (the light reactions, sucrose production etc.) and so 
we have to make a choice of the set of chemical species whose concentrations 
are included as unknowns in the ODE system. The hope is that these 
concentrations have only a small effect on the concentrations of the other 
species with which the chosen ones interact. The concentrations of these 
other species are taken to be constant and we refer to them as external species
while the species whose concentrations are the unknowns in the ODE system are 
referred to as internal species. A possible justification for this procedure 
is that if the concentration of an external species is very high it will remain 
approximately constant even if some amount of the substance concerned is being 
produced or consumed by some of the other reactions. There is also a choice of 
which reactions are considered to be taking place at an appreciable rate. 
Usually the stoichiometry of the reactions is known but the same cannot be 
said of the reaction rates. There further assumptions have to be made. Summing 
up, different mathematical models arise through different choices of the 
species and reactions included and the reaction kinetics.
Furthermore it may happen that models are replaced by smaller ones using 
limits involving time scale separation or elimination of intermediate species
in some reactions.

After these preliminary considerations we may look at what a standard textbook
on cell biology \cite{alberts02} tells us about the Calvin cycle. The essential 
features of this process were worked out by Melvin Calvin and his collaborators
(earning Calvin the Nobel prize for chemistry in 1961). Often the situation of 
carbon dioxide and light saturation is considered. Calvin's experiments were
done under these circumstances and they are often assumed to hold when doing 
modelling. This means on the one hand that carbon dioxide is so plentiful that 
it can be considered as an external species. On the other hand the substances 
ATP and NADPH which are supplied by the light reactions are assumed to be 
plentiful. Thus CO${}_2$, ATP and NADPH are taken as external species. The 
same is true of ADP and NADP which are produced from ATP and NADPH in certain 
reactions. Inorganic phosphate ${\rm P_i}$ is often also treated as an external
species. All the reactions are catalysed by enzymes but these are usually 
treated as external species. The substances which remain in the description in 
\cite{alberts02}, and which will be internal species, are 
ribulose 1,5-bisphosphate (RuBP), 3-phosphoglycerate (PGA), 
1,3-bisphosphoglycerate (DPGA), glyceraldehyde 3-phosphate (GAP) and ribulose 
5-phosphate (Ru5P). The simplest assumption is that each of these substances
reacts to give the next with a final reaction taking us back from Ru5P to RuBP.
Thus we have a cycle, explaining the other part of the name 'Calvin cycle'.
While most of these are bona fide reactions, that leading from GAP to Ru5P is
an effective reaction (or, less respectfully, a fudge reaction)
resulting from collapsing part of a more complicated network. This can easily 
be recognized by means of the exotic stoichiometry, with 
five molecules going in and three coming out. In addition there are two 
transport processes in which PGA and GAP are exported to the cytosol from the 
chloroplast where the Calvin cycle takes place. Thus some simple models of
photosynthesis (to be considered in more detail in the next section) have
five species and seven reactions.

In this paper there is no attempt to present a systematic catalogue of models.
Instead it is like an accompanied  walk through a zoo, where the visitor is
taken to see the lions and the elephants but also less familiar exhibits such 
as the giant anteater or the Tasmanian devil. It starts with the simplest and 
best known models and is led by the consideration of various issues to ones 
which have been studied less. 

So many related models are considered in what follows that it would be 
cumbersome to have a name for each of them. Some names will be used but in
addition the models will be given numbers according to the pattern Model m.n.k,
where roughly speaking this means the variant $k$ of the model $n$ first
introduced in section $m$.  

The paper is organized as follows. Section \ref{5species} introduces the 
models with five species and is mainly a survey of known results concerning 
them. These provide information about the existence and stability of positive 
steady states and the existence of solutions where the concentrations tend to 
zero or to infinity at late times. The process of passing from one model to 
another by making an internal species into an external one is carried out in a 
simple example. In Section \ref{pettersson} models with a larger number of
variables (about fifteen) are considered which are variants of one introduced
in \cite{pettersson88}. Some known results on the (non)-existence of positive
steady states and the ways in which concentrations can approach zero at late
times are reviewed and extended. It is shown how the model of 
\cite{pettersson88} itself can be given a clear mathematical formulation as a 
system of DAE. It is also shown how ATP can be made into an external species 
in these models. Section \ref{steadype} contains a proof that the model of 
\cite{pettersson88} admits more than one positive steady state for suitable 
values of the parameters. This is related to a stoichiometric generator for 
the network and other generators, which might also be helpful in the search for 
steady states, are presented. A similar approach can be applied to the related
reduced Poolman model and this is done in Section \ref{steadypo}, where it is
shown that there are parameter values for which there exist at least three
positive steady states. The last section is concerned with some prospects for 
future progress and briefly discusses some simplified models due to Hahn. 

\section{The five-species models}\label{5species}

In this section some models will be considered where the unknowns are
the concentrations of the five substrates introduced in the previous section. 
The concentration of a substance $X$ is denoted by $x_X$. The reactions are 
those mentioned in the introduction and they are all assumed to be 
irreversible. The simplest type of kinetics which 
can be assumed is mass action kinetics. For this we need to know the exact 
stoichiometry. At this point there is an ambiguity resulting from the fudge 
reaction. In that case the biochemistry does not determine unique values for
the stoichiometric coefficients. It only determines the ratio of the number
of molecules going into the reaction and the number coming out. One possible
choice, which was made in \cite{zhu09}, is to use the coefficients $1$ and 
$0.6$. Since this is only an effective reaction there is no strong argument
that these coefficients should be integers. Nevertheless in \cite{grimbs11}
the authors preferred to use $5$ and $3$. This does make a difference to the
evolution equations resulting from the assumption of mass action kinetics.
With the stoichiometry of \cite{grimbs11} mass action kinetics leads to 
nonlinear evolution equations (Model 2.1.1) while the kinetics of \cite{zhu09}
gives linear equations (Model 2.1.2). An alternative to mass action kinetics
is Michaelis-Menten kinetics, either with the coefficients of \cite{grimbs11}
(Model 2.2.1) or those of \cite{zhu09} (Model 2.2.2).

We now examine the dynamics of these models. It was shown in \cite{grimbs11}
that if the reaction constants satisfy a certain inequality ($k_2\le 5k_6$) 
Model 2.1.1 does not possess any positive steady states while if $k_2>5k_6$ it 
possesses precisely one positive steady state for given values of the 
parameters. These statements are obtained by explicit calculation.
In the latter case it was shown that the steady state is unstable.
This could be seen as a disappointment since it might be supposed for 
biological reasons that the cycle can exist in a stable configuration. This is 
not absolutely clear since a good model need not give a good description of 
the dynamics globally in time but only on a time scale long enough so as to 
capture the processes which are to be described. This quantitative line of 
thought will not be pursued further here. In \cite{rendall14} it was further 
shown that in the case $k_2\le 5k_6$ all concentrations tend to zero as time 
tends to infinity. This was done with the help of a Lyapunov function. In 
\cite{grimbs11} and \cite{rendall14} the inequality relating $k_2$ and $k_6$ 
was not given any biological interpretation and no explanation was given for 
the Lyapunov function which was found by trial and error. More insight on 
these questions was obtained while studying a more complicated model of the 
Calvin cycle in \cite{moehring15}. The reaction constant $k_6$ controls the 
rate of export of PGA from the chloroplast and in reality this is coupled to 
the import of inorganic phosphate. Thus intuitively $k_6$ secretly contains a 
dependence on the constant concentration of inorganic phosphate in
the cytosol. The positive steady state disappears when there is too much
phosphate in the cytosol. Then the production of sugars by the cycle cannot
keep up with the export. This phenomenon has been called overload breakdown
\cite{pettersson88}. To obtain some insight into the Lyapunov function it is
helpful to consider the total number of carbon atoms in the system. The 
reactions within the cycle conserve carbon apart from the fact that at one point
carbon dioxide is imported. The export processes also do not conserve carbon.
Nevertheless the time derivative of the total amount of carbon only has a few 
contributions. One of them has a positive sign but modifying the coefficient of
$x_{\rm PGA}$ allows a Lyapunov function $L_1$ to be obtained. There remains the 
question of what happens to solutions of Model 2.1.1 at late times when 
$k_2>5k_6$. It was shown in \cite{rendall14} that there are solutions where 
all concentrations tend to zero at late times (using a modification $L_2$ of
$L_1$) and solutions for which the concentrations tend to infinity as 
$t\to\infty$ (runaway solutions) and their leading order asymptotics were 
determined. From what has just been said it can be seen that 
a number of facts are known about the dynamics of Model 2.1.1 but there remain
open questions, for instance whether periodic solutions exist. Information
about Model 2.1.2 has also been obtained in \cite{rendall14}. In particular, 
there are either no positive steady states or a whole continuum of steady 
states, depending on the values of the reaction constants. Both solutions 
converging to the origin at late times and runaway solutions occur.

When mass action kinetics is replaced by Michaelis-Menten kinetics there are
still runaway solutions but there are also more interesting steady states. 
Concerning Model 2.2.2 it stated in \cite{zhu09} that there is at most one 
steady state which is \lq physiologically feasible\rq. This last condition 
includes 
restrictions on the values of the model parameters. In some cases a parameter 
interval is chosen centred at a value taken from the experimental literature. 
Some Michaelis constants $K_{mi}$ are set to fixed values and it is not clear 
to this author where these values come from. The paper \cite{zhu09} uses 
computer-assisted methods which are claimed to prove the assertion about the
limitation on steady states. In \cite{disselnkoetter16} a purely analytical 
proof of the assertion was given under the assumptions on the Michaelis 
constants made in \cite{zhu09}. It was also proved that the assertion depends 
essentially on these assumptions. There are examples with 
$\kappa=(K_{m7}-K_{m4})(K_{m6}-K_{m21})<0$ for which there exist two distinct 
positive steady states and if both factors in the expression for $\kappa$ 
vanish there is a continuum of steady states. It was also shown that there 
exist cases with two isolated steady states where one of them is stable and 
the other is unstable. This is proved by showing that there is a bifurcation 
with a one-dimensional centre manifold. In particular we obtain models admitting
a stable positive steady state although it is unclear whether the parameter
values required for this are biologically reasonable. Model 2.2.1 also 
permits the existence of two positive steady states, one of which is stable
and the other is unstable.

Another type of model, considered in \cite{grimbs11}, is obtained if each of 
the basic reactions considered up
to now is replaced by a Michaelis-Menten scheme with a substrate, an enzyme 
and a substrate-enzyme complex (Models 2.3.1 and 2.3.2) and these will be 
discussed in this section although they contain many more than five variables.
It is possible to pass from these models to Models 2.2.1 and 2.2.2 by a
Michaelis-Menten reduction which is well-behaved in the sense of geometric
singular perturbation theory (GSPT) - the transverse eigenvalues have negative 
real parts. (For background information on GSPT we refer to \cite{kuehn15}.) 
This means that we can transfer information on the existence and stability of
steady states from Models 2.2.1 and 2.2.2 to Models 2.3.1 and 2.3.2 in a 
straightforward way. In fact the existence of more than one steady state of
Model 2.3.1 was discovered directly in \cite{grimbs11} with the help of
elementary flux modes.

In another model introduced in \cite{grimbs11} ATP was made into an internal 
species, thus producing a six-variable model and diffusion of ATP was included.
Restricting consideration to spatially homogeneous solutions reduces the 
resulting system of reaction-diffusion equations to a system of ODE (Model
2.4.1). It turns out that Model 2.4.1 can be analysed as in the cases of 
Models 2.2.1 and 2.2.2, giving the existence of two steady states, one stable
and one unstable \cite{disselnkoetter16}. Interestingly, the solutions of
Model 2.4.1 are bounded although this is non-trivial to prove \cite{rendall14}.
In all these models $\omega$-limit points where some concentration vanishes
are strongly restricted. In Models 2.1.1, 2.1.2, 2.2.1 and 2.2.2 the only point 
which can occur is the origin. In Model 2.3.1 the analogue of this is a 
situation where all substrates are exhausted and the enzymes are completely
in the unbound form. In Model 2.4.1 the corresponding situation is that all
concentrations except that of ATP are zero and the concentration of ATP takes
on its maximal value.

The process of making a species into an external species can be illustrated
by showing how Model 2.1.1 can be obtained as a limit of Model 2.4.1. The
evolution equations of Model 2.4.1 are
\begin{eqnarray}
&&\frac{dx_{\rm RuBP}}{dt}=k_5x_{\rm Ru5P}x_{\rm ATP}-k_1x_{\rm RuBP},\label{madh1}\\
&&\frac{dx_{\rm PGA}}{dt}=2k_1x_{\rm RuBP}-k_2x_{\rm PGA}x_{\rm ATP}
-k_6x_{\rm PGA},
\label{madh2}\\
&&\frac{dx_{\rm DPGA}}{dt}=k_2x_{\rm PGA}x_{\rm ATP}-k_3x_{\rm DPGA},\label{madh3}\\
&&\frac{dx_{\rm GAP}}{dt}=k_3x_{\rm DPGA}-5k_4x_{\rm GAP}^5-k_7x_{\rm GAP},
\label{madh4}\\
&&\frac{dx_{\rm Ru5P}}{dt}=-k_5x_{\rm Ru5P}x_{\rm ATP}
+3k_4x_{\rm GAP}^5,\label{madh5}\\
&&\frac{dx_{\rm ATP}}{dt}=-k_2x_{\rm PGA}x_{\rm ATP}-k_5x_{\rm Ru5P}x_{\rm ATP}
+k_8(c_A-x_{\rm ATP})\label{madh6}
\end{eqnarray}
where the constant $c_A$ is the total concentration of adenosine phosphates.
We would now like to consider a situation where ATP is in excess. Equivalently
we can consider a situation where the concentrations of all substances except
ATP are very small. Define $x_X=\eta\tilde x_X$ for each substance X 
except ATP. Define $\tilde k_4=\eta^4k_4$. Then making these substitutions
gives 
\begin{eqnarray}
&&\frac{d\tilde x_{\rm RuBP}}{dt}=k_5\tilde x_{\rm Ru5P}x_{\rm ATP}
-k_1\tilde x_{\rm RuBP},\label{madht1}\\
&&\frac{d\tilde x_{\rm PGA}}{dt}=2k_1\tilde x_{\rm RuBP}
-k_2\tilde x_{\rm PGA}x_{\rm ATP}
-k_6\tilde x_{\rm PGA},
\label{madht2}\\
&&\frac{d\tilde x_{\rm DPGA}}{dt}=k_2\tilde x_{\rm PGA}x_{\rm ATP}
-k_3\tilde x_{\rm DPGA},\label{madht3}\\
&&\frac{d\tilde x_{\rm GAP}}{dt}=k_3\tilde x_{\rm DPGA}-5\tilde k_4\tilde x_{\rm GAP}^5
-k_7\tilde x_{\rm GAP},
\label{madht4}\\
&&\frac{d\tilde x_{\rm Ru5P}}{dt}=-k_5\tilde x_{\rm Ru5P}x_{\rm ATP}
+3\tilde k_4\tilde x_{\rm GAP}^5,\label{madht5}\\
&&\frac{d\tilde x_{\rm ATP}}{dt}=-\eta k_2\tilde x_{\rm PGA}x_{\rm ATP}
-\eta k_5\tilde x_{\rm Ru5P}x_{\rm ATP}
+k_8(c_A-x_{\rm ATP}).\label{madht6}
\end{eqnarray}
Letting $\eta$ tend to zero gives a system for which $x_{\rm ATP}=c_A$ is an 
invariant manifold and the restriction of the system to that manifold 
reproduces the equations of Model 2.1.1. This is a regular limit and it 
follows from the existence of an unstable hyperbolic positive steady state in 
the limiting system that Model 2.4.1 also has an unstable hyperbolic positive 
steady state. Note that the perturbed steady state for $\eta$ small and 
positive does satisfy $x_{\rm ATP}<c_A$ since otherwise equation
(\ref{madht6}) would lead to a contradiction. For this system the information
about a steady state obtained by the perturbation argument is less than 
what is already known by analysing the full system directly. The argument
has nevertheless been presented here since analogous arguments may be useful
for obtaining information about more complicated systems where no alternative
is available.

\section{The Pettersson model and modifications of it}
\label{pettersson}

The models considered in this section involve more unknowns than those of the 
previous section. The starting point is a model introduced in a paper by 
Pettersson and Ryde-Pettersson \cite{pettersson88} which we refer to for
brevity as the Pettersson model. The substances included are roughly speaking 
those which Calvin found to appear after the dark reactions had run for a few 
minutes. In addition to those in the five-variable models these are 
dihydroxyacetone phosphate (DHAP), fructose 
1,6-bisphosphate (FBP), fructose 6-phosphate (F6P), erythrose 4-phosphate 
(E4P), sedoheptulose 7-phosphate (S7P), sedoheptulose 1,7-bisphosphate (SBP), 
xylulose 5-phosphate (X5P) and ribose 5-phosphate (R5P). In addition the 
process by which sugars can be stored in the chloroplast as starch is 
included. Starch itself is treated as an external species. The intermediates 
glucose 6-phosphate (G6P) and glucose 1-phosphate (G1P) are included as 
internal species. In contrast to the models of the previous section inorganic 
phosphate in the chloroplast, $\rm P_i$,
is modelled dynamically, as is ATP. On the other hand NADPH is still 
treated as an external species. Some of the reactions which were treated as 
irreversible in the models of the previous section are treated as reversible
in the Pettersson model, for instance the reaction interconverting PGA and 
DPGA. The decision, which reactions to treat as reversible and which as 
irreversible in the Pettersson model is based on experimental data. The 
only reactions treated as irreversible are those whose substrates are Ru5P,
RuBP, FBP and G1P together with the transport processes to the cytosol and
to starch.  

The Pettersson model (Modell 3.1.1) will now be described. In 
\cite{pettersson88} the time derivatives of the relevant concentrations are 
expressed in terms of the rates $v_i$ of the different reactions. The 
equations are
\begin{eqnarray}
&&\frac{d x_{\rm RuBP}}{dt}=v_{13}-v_1,\\
&&\frac{d x_{\rm PGA}}{dt}=2v_1-v_2-v_{\rm PGA},\\
&&\frac{d x_{\rm DPGA}}{dt}=v_2-v_3,\\ 
&&\frac{d x_{\rm ATP}}{dt}=v_{16}-v_2-v_{13}-v_{\rm st},\\
&&\frac{d x_{\rm GAP}}{dt}=v_3-v_4-v_5-v_7-v_{10}-v_{\rm GAP},\\
&&\frac{d x_{\rm DHAP}}{dt}=v_4-v_5-v_8-v_{\rm DHAP},\\
&&\frac{d x_{\rm FBP}}{dt}=v_5-v_6,\\
&&\frac{d x_{\rm F6P}}{dt}=v_6-v_7-v_{14},\\
&&\frac{d x_{\rm E4P}}{dt}=v_7-v_8,\\
&&\frac{d x_{\rm X5P}}{dt}=v_7+v_{10}-v_{12},\\
&&\frac{d x_{\rm SBP}}{dt}=v_8-v_9,\\
&&\frac{d x_{\rm S7P}}{dt}=v_9-v_{10},\\
&&\frac{d x_{\rm R5P}}{dt}=v_{10}-v_{11},\\
&&\frac{d x_{\rm Ru5P}}{dt}=v_{11}+v_{12}-v_{13},\\
&&\frac{d x_{\rm G6P}}{dt}=v_{14}-v_{15},\\
&&\frac{d x_{\rm G1P}}{dt}=v_{15}-v_{\rm st},\\
&&\frac{d x_{\rm P_i}}{dt}=v_3+v_6+v_9+v_{\rm PGA}+v_{\rm GAP}+v_{\rm DHAP}
+2v_{\rm st}-v_{16}.
\end{eqnarray}
The total amount of phosphate in the chloroplast is a conserved quantity and 
may be used to eliminate the concentration of inorganic phosphate in the 
chloroplast from the equations in favour of the other variables.
It is assumed that the reversible reactions are much faster than the 
irreversible ones. This can be implemented mathematically by introducing
a small parameter $\epsilon$ and defining $\tilde v_i=\epsilon v_i$ for the 
fast reactions. The slow reactions, whose rates are not rescaled, are those 
with reaction rates $v_1$, $v_6$, $v_9$, $v_{13}$, $v_{16}$, $v_{\rm PGA}$, 
$v_{\rm GAP}$, $v_{\rm DHAP}$ and $v_{\rm st}$. For each of the slow reactions an 
explicit phenomenological expression is given for the rate. This incorporates 
the known experimental facts on the activation and inhibition of certain 
reactions due to the influence of other substances. No expressions are given 
for the rates of the fast reactions. Instead it is assumed that these 
reactions can be taken to be in equilibrium, which gives algebraic equations 
relating the concentrations. It will be shown below how this can be 
implemented mathematically. 

At this point we interrupt the discussion of the Pettersson model and instead 
take the reaction network underlying the Pettersson model including its 
stoichiometry and apply mass action kinetics to get something which was called 
the Pettersson-MA model in \cite{moehring15} (Model 3.2.1). The strategy 
adopted in \cite{moehring15} was to study the dynamics of Model 3.2.1 so as to 
try to obtain some insights for tackling the Pettersson model later. There is 
a related model with an additional reaction which liberates G1P from starch 
(Model 3.2.2). There the above evolution equations are modified by adding
a contribution $v_{17}$ to the evolution for $x_{\rm G1P}$ and a contribution
$-v_{17}$ to the evolution equation for $x_{\rm P_i}$. 
This will be called the Poolman-MA model since a modification 
of the Pettersson model including a mechanism of this type was first 
introduced by Poolman \cite{poolman99}. Poolman himself used the same reaction
rates as in \cite{pettersson88} for the slow reactions and treated the 
liberation of G1P as a slow reaction while taking mass action kinetics for the
fast reactions. We call the resulting system of ODE the Poolman model 
(Model 3.1.2). A hybrid model can be obtained by taking the reactions 
included in the Pettersson model with the reaction rates as in the Poolman 
model (Model 3.3.1). This is obtained from the Poolman model by setting one of 
the reaction constants $k_{32}$ to zero. In these models the total amount of 
phosphate is conserved and every unknown contains some phosphate. Thus the 
conservation law implies that all solutions are bounded and runaway solutions 
are ruled out for these models. It was already mentioned that the export of 
sugars from the chloroplast is coupled to the import of inorganic phosphate 
(whose concentration in the cytosol is assumed constant in the model). It is 
indicated in \cite{pettersson88} that if the external concentration of 
phosphate is too high then no positive steady state will exist. This is the 
phenomenon of overload breakdown. Poolman suggested that overload breakdown 
could be avoided by introducing the release of G1P from starch. In the case of 
Model 3.2.1 it was shown in \cite{moehring15} that if $k_3c_A\le 5k_{28}$ there 
exists a Lyapunov function related to the function $L_1$ of the last section 
and this proves that under this condition Model 3.2.1 has no positive steady 
states. Here the $k_i$ are reaction constants and $c_A$ is the total 
concentration of adenosine phosphates. For Model 3.2.2 this construction no 
longer works. When $k_3c_A>5k_{28}$ in Model 3.2.1 it is possible to construct 
an analogue of the function $L_2$ of the last section which gives the 
conclusion that there exist no positive steady states where $L_2$ is less than 
a certain number depending only on the reaction constants. Here we define
\begin{equation}
L_2=L_1-\frac12 (x_{\rm DPGA}+x_{\rm GAP}+x_{\rm DHAP}).
\end{equation}
and it satisfies 
\begin{eqnarray}
&&\frac{d(5L_2)}{dt}=\frac12 \left(2k_9x_{\rm DHAP}x_{\rm GAP}
+k_{12}x_{\rm FBP}x_{\rm GAP}+k_{14}x_{\rm E4P}x_{\rm DHAP}\right.\nonumber\\
&&\left.+k_{17}x_{\rm S7P}x_{\rm GAP}
-5k_{29}x_{\rm GAP}-5k_{30}x_{\rm DHAP}-k_{11}x_{\rm E4P}x_{\rm X5P}\right.\nonumber\\
&&\left.-k_{16}x_{\rm X5P}x_{\rm R5P}-5x_{\rm PGA}-k_8x_{\rm FBP}-k_{13}x_{\rm SBP}\right).
\end{eqnarray}
When $L_2$ is sufficiently small the positive terms on the right hand side are
dominated by the negative ones. Information can also be obtained
for Model 3.3.1 by using the function $L_1$. In that case $L_1$ is decreasing
provided the quantity $\frac12 k_3x_{\rm ATP}-\frac52 v_{\rm PGA}$ is negative. Now
$v_{\rm PGA}=\frac{V_{\rm ex}x_{\rm PGA}}{NK_{\rm PGA}}$ where
\begin{equation}
N=1+\left(1+\frac{K_{\rm P_{ext}}}{x_{\rm P_{ext}}}\right)
\left(\frac{x_{\rm P_i}}{K_{\rm P_i}}+\frac{x_{\rm PGA}}{K_{\rm PGA}}
+\frac{x_{\rm PGA}}{K_{\rm GPA}}+\frac{x_{\rm DHAP}}{K_{\rm DHAP}}\right)
\end{equation}
and the other quantities which have not previously been defined are positive 
constants. Here we treat the total amount of phosphate as a parameter and 
then if $k_3$ is chosen small enough for fixed values of the other parameters 
in the kinetics we get a positive lower bound for $x_{\rm PGA}^{-1}v_{\rm PGA}$. 
Thus under these conditions $L_1$ is decreasing. In particular Model 3.3.1 
has no positive steady states when the parameters are restricted in this way.

In \cite{moehring15} conditions were derived for $\omega$-limit points of 
positive solutions of Models 3.2.1 and 3.2.2. It was pointed out in 
\cite{moehring15} that many of the arguments used apply to the original
Poolman model since the only property of the reaction rates which is used
is under what circumstances they are positive or zero and this is not
changed when the mass action kinetics is replaced by the more complicated
kinetics of the Poolman model. The same argument applies to the hybrid
model. It thus follows from the arguments in \cite{moehring15} that 
Models 3.1.2, 3.2.1, 3.2.2 and 3.3.1 have the property that the only
substances whose concentrations may fail to vanish at an $\omega$-limit 
point of a positive solution where at least one concentration vanishes are 
G1P, G6P, F6P, E4P, S7P and $\rm P_i$. 

In \cite{moehring15} information was also obtained on how these points may be 
approached by positive solutions of Models 3.2.1 and 3.2.2. This is done by
linearizing about steady states where some concentrations are zero and 
analysing the eigenvalues of the linearization. In some cases spectral 
stability could be determined but other cases remain open. In many cases where
the spectral analysis was successful it turned out that the centre manifold 
coincides with the center subspace and the qualitative behaviour on the centre
manifold could be analysed. In other cases the centre manifold is nonlinear 
and its Taylor expansion not yet been computed.

Consider now again the evolution equations for the Pettersson model expressed
in terms of the reaction rates $v_i$. In \cite{pettersson88} five linear 
combinations $y_i$ of concentrations are identified whose time derivatives 
only depend on the slow reaction rates. Suppose we now complement these by a 
suitable set of concentrations $z_i$, for instance all those except 
$x_{\rm RuBP}$, $x_{\rm F6P}$, $x_{\rm Ru5P}$, $x_{\rm DHAP}$ and $x_{\rm ATP}$, which we 
denote by $s_i$. Then the concentrations of the internal species are related 
to the variables $y_i$ and $z_i$ by an invertible linear transformation. 
Consider the equations of the hybrid model, expressed in terms of the 
variables $y_i$ and $z_i$. If we write them in terms of the $\tilde v_i$ then 
all the terms on the right hand side of the evolution equations for the $y_i$ 
are regular in $\epsilon$ while many of those on the right hand side of the 
equations for the $z_i$ contain a factor $\epsilon^{-1}$. Multiplying these
equations with $\epsilon$ and letting $\epsilon$ tend to zero gives a 
system of algebraic equations. When expressed in terms of the $\tilde v_i$ 
these equations are linear and they imply that the $\tilde v_i$ vanish for all 
the fast reactions at $\epsilon=0$. This fact can be read off from the 
subnetwork obtained by deleting the slow reactions from the full Pettersson
network. Now the $\tilde v_i$ can be obtained from 
$v_i$ by replacing the reaction constants $k_i$ by $\tilde k_i=\epsilon k_i$. If 
the $\tilde k_i$ are chosen independent of $\epsilon$ the algebraic equations 
(20)-(30) of \cite{pettersson88} for the concentrations are obtained. Thus in 
the limit $\epsilon\to 0$ the hybrid model becomes a system consisting of the
differential equations (48) and the algebraic equations (20)-(30) of 
\cite{pettersson88}. It defines a system of DAE for the variables $y_i$ and 
$z_i$. This is what we refer to as the Pettersson model. Without 
further information it is not even clear that local existence holds for this 
system. In \cite{pettersson88} it is claimed that the equations (20)-(30) of 
that paper can be used to obtain an explicit closed system of evolution 
equations for the variables $s_i$ but the calculations presented there are not 
complete. A similar reduction to a DAE can be carried out for the Poolman
model and we call the result the reduced Poolman model (Model 3.3.2).

Any of the models considered in this section may be modified so as to make 
ATP and inorganic phosphate external species. By analogy with what was done
in the case of Model 2.4.1 we can fix the concentration of ATP and $c_A$ and 
rescale the concentrations of the other substances by a factor $\eta$. The 
fact of having eliminated the concentration of inorganic phosphate by using 
the conservation of the total amount of phosphate means that $x_{\rm P_i}$ then
automatically becomes constant. In the same way as $k_4$ had to be scaled by 
a power of $\eta$ in Model 2.4.1 it is necessary to rescale the reaction
constants in the reactions with two substrates in order to get a non-zero
limit. For instance, using the notation of \cite{moehring15} we introduce 
$\tilde k_9=\eta k_9$. The other reaction constants which should be 
rescaled in a similar manner are $k_{11}$, $k_{12}$, $k_{14}$, $k_{16}$ and $k_{17}$.
To maintain the non-trivial effects of saturation, activation and inhibition 
in the slow reactions it is also nessary to rescale certain Michaelis 
constants. The constants involved are $K_{m6}$, $K_{i61}$, $K_{m9}$, $K_{mst1}$, 
$K_{m1}$, $K_{i11}$, $K_{i12}$, $K_{i13}$, $K_{i15}$, $K_{m131}$, $K_{i131}$, 
$K_{i132}$, $K_{ast1}$, $K_{ast2}$, $K_{ast3}$, $K_{\rm PGA}$, $K_{\rm GAP}$ and 
$K_{\rm DHAP}$. Let us call the results of modifying Models 3.1.1 and 3.3.1 in 
this way Models 3.4.1 and 3.4.2 respectively. In both of these cases the 
modified model has an invariant manifold $x_{\rm ATP}=c_A$ for $\eta=0$ and 
the restriction of the system to that submanifold reproduces the model with 
ATP as an external species. As in the discussion of Model 2.4.1 in the last 
section this means that any hyperbolic positive steady state of Model 3.4.1 or 
3.4.2 gives rise to a hyperbolic positive steady state of Model 3.1.1 or 
3.3.1, respectively. Thus information on steady states can be obtained from 
information on the corresponding models with the concentrations of ATP and 
$\rm P_i$ frozen.
   
\section{Steady states of the Pettersson model}\label{steadype}

Five of the equations in the Pettersson model are evolution equations. The 
right hand sides of these equations are the functions $F_i$ in the equations 
(54)-(58) of \cite{pettersson88} and their vanishing is equivalent to the 
equations (42)-(47) of \cite{pettersson88}, which are, with the definition 
$v=v_1$,
\begin{eqnarray}
&&v_1=v,\label{Fi01}\\
&&v_6=\frac{v}{3}+v_{\rm st},\label{Fi02}\\
&&v_9=\frac{v}{3},\label{Fi03}\\
&&v_{13}=v\label{Fi04},\\
&&v_{16}=3v+v_{\rm st}-v_{\rm PGA},\label{Fi05}\\
&&v=3v_{\rm ex}+6v_{\rm st}\label{Fi06}
\end{eqnarray} 
where $v_{\rm ex}=v_{\rm PGA}+v_{\rm GAP}+v_{\rm DHAP}$.
In this section we concentrate on Model 3.4.1, where all these equations hold
except that containing $v_{16}$. Suppose that $x_{\rm DHAP}$ is given. Then 
$\tilde v_4=0$ fixes the value of $x_{\rm GAP}$. Then $\tilde v_3=0$ and 
$\tilde v_2=0$ fix the values of $x_{\rm DPGA}$ and $x_{\rm PGA}$. In addition 
$\tilde v_5=0$ fixes the value of $x_{\rm FBP}$. With the information we have
it is possible to compute $v_{\rm ext}$ in terms of $x_{\rm DHAP}$. Suppose now 
that $v_{\rm st}$ is also fixed. Then with this information it is possible to 
obtain $v$ and hence $v_9$ and $v_{13}$. The quantities $x_{\rm SBP}$, $x_{\rm RuBP}$ 
and $x_{\rm Ru5P}$ are then uniquely determined. To ensure the existence of
these quantities it suffices to assume that the parameters $V_9$, $V_{13}$
and $V_1$ are large enough. The equations 
$\tilde v_{11}=0$ and $\tilde v_{12}=0$ fix $x_{\rm R5P}$ and $x_{\rm X5P}$. Next 
the equation $\tilde v_8=0$ allows $x_{\rm E4P}$ to be determined. Then 
$\tilde v_7=0$ can be used to determine $x_{\rm F6P}$ and $\tilde v_{14}=0$ and 
$\tilde v_{15}=0$ give $x_{\rm G6P}$ and $x_{\rm G1P}$. This leaves two consistency 
conditions, namely the equation for $v_6$ and the expression for $v_{\rm st}$ in 
terms of $x_{\rm G1P}$. Let these be denoted abstractly by 
$\Phi_1(x_{\rm DHAP},v_{\rm st})=0$ and $\Phi_2(x_{\rm DHAP},v_{\rm st})=0$
respectively.

A general strategy for looking for positive steady states of the evolution
equations defined by a given network is to look at limiting cases where some 
of the reaction constants are set to zero and the smaller network obtained
by discarding the reactions concerned. Trying to do this for a larger network
without some guiding principles may fail because there are too many 
possibilities. A concept which can be used as a guiding principle is that
of elementary flux modes. There is a theory of how to produce steady states
using these objects \cite{conradi07} but an alternative is to use elementary 
flux mode to guess which reaction constants to set to zero and then proceed 
directly with the construction of steady states. This possibility was used in
\cite{rendall14} to give an existence proof of steady states of Model 2.3.1 
and it will also be applied in what follows. The elementary flux modes 
computed are not required for the proofs themselves but they help to put those 
proofs into context. They helped to find the proofs and this approach might 
also turn out to be useful for analysing other similar models in the future.

Consider the equations satisfied by the fluxes in steady states of a model 
defined by the Pettersson network. These are
\begin{eqnarray}
&&2v_1-v_2-v_{\rm PGA}=0,\label{fluxb1}\\
&&v_2=v_3,\label{fluxb2}\\
&&v_3-v_4-v_5-v_7-v_{10}-v_{\rm GAP}=0,\label{fluxb3}\\
&&v_4-v_5-v_8-v_{\rm DHAP},\label{fluxb4}\\
&&v_5=v_6,\label{fluxb5}\\
&&v_6-v_7-v_{14}=0,\label{fluxb6}\\
&&v_7=v_8,\label{fluxb7}\\
&&v_8=v_9,\label{fluxb8}\\
&&v_9=v_{10},\label{fluxb9}\\
&&v_7+v_{10}-v_{12}=0,\label{fluxb10}\\
&&v_{10}=v_{11},\label{fluxb11}\\
&&v_{11}+v_{12}-v_{13}=0,\label{fluxb12}\\
&&v_{13}=v_1,\label{fluxb13}\\
&&v_{14}=v_{15},\label{fluxb14}\\
&&v_{15}=v_{\rm st},\label{fluxb15}\\
&&v_{16}-v_2-v_{13}-v_{\rm st}=0.\label{fluxb16}
\end{eqnarray}
The solutions of these linear equations can be parametrized with the help of 
stoichiometric generators. The relevant terminology  will now be recalled 
(cf. \cite{conradi07}). The system of ODE arising from a reaction network can 
be written in the form $\dot x=Nv(x)$, where $N$ is the stoichiometric matrix 
and $v(x)$ are the reaction rates. In this context reversible reactions are 
treated as two separate reactions. This means that there are two columns of 
$N$ corresponding to each reversible reaction. The column corresponding to the 
forward reaction is minus the column corresponding to the backward reaction. 
If we discard one of the two columns corresponding to each reversible reaction 
a truncated matrix $\bar N$ is obtained. The kernels of the matrices $N$ and 
$\bar N$ are related in a simple way which will be described below. The set of 
reaction rates at a steady state is an element of the kernel of $N$ with 
non-negative components. We can think of this as a point in the space of 
real-valued functions on the set $\cal R$ of reactions. The set of all 
non-negative elements of the kernel of $N$ is a positive cone and thus 
consists of all vectors of the form  $\sum_i a_iw_i$ with $a_i$ non-negative 
coefficients and $w_i$ a finite number of vectors which in this context are 
called elementary flux modes \cite{rockafellar70}, \cite{conradi07}. Each of 
these vectors has the property that
setting some but not all of its components to zero gives a vector which is 
not in the kernel of $N$. Another important quantity is the incidence
matrix. It has one row for each complex (quantity on the left or right hand 
side of a reaction) and one column for each reaction. The element for the
left hand side of the reaction is $-1$, the element for the right hand side is
$+1$ and all other elements are zero. A vector which is in the kernel of the
incidence matrix is in the kernel of $N$. Elementary flux modes which are
not in the kernel of the incidence matrix are called stoichiometric 
generators. For a reversible reaction let us make a choice of which is the 
forward direction, so as to get a forward reaction $r_+$ and a backward 
reaction $r_-$. Then the vector which has the components $+1$ at $r_+$, $-1$ 
at $r_-$ and all other components zero belongs to the kernel of the incidence 
matrix. Let us call this a trivial mode. It is an elementary flux mode which 
is not a stoichiometric generator. The kernel of $N$ is the joint span of the
kernel of $\bar N$ and the trivial modes.  

In the case of the Pettersson network the following vectors with components 
$w_i$ are stoichiometric generators
\begin{eqnarray}
&&[3\ 6\ 6\ 3\ 1\ 1\ 1\ 1\ 1\ 1\ 1\ 2\ 3\ 0\ 0\ 9\ 0\ 0\ 1\ 0],\label{sg1}\\
&&[3\ 6\ 6\ 2\ 1\ 1\ 1\ 1\ 1\ 1\ 1\ 2\ 3\ 0\ 0\ 9\ 0\ 1\ 0\ 0],\label{sg2}\\
&&[3\ 5\ 5\ 2\ 1\ 1\ 1\ 1\ 1\ 1\ 1\ 2\ 3\ 0\ 0\ 8\ 1\ 0\ 0\ 0],\label{sg3}\\
&&[6\ 12\ 12\ 5\ 3\ 3\ 2\ 2\ 2\ 2\ 2\ 4\ 6\ 1\ 1\ 19\ 0\ 0\ 0\ 1].\label{sg4}
\end{eqnarray} 
Here the components are written in the order 
\begin{equation}
[v_1\ v_2\ v_3\ v_4\ v_5\ v_6\ v_7\ v_8\ v_9\ v_{10}\ v_{11}\ v_{12}\ v_{13}
\ v_{14}\ v_{15}\ v_{16}\ v_{\rm PGA}\ v_{\rm GAP}\ v_{\rm DHAP}\ v_{\rm st}].
\end{equation}
We only write out the components corresponding to the forward reactions.
In other words these vectors belong to the kernel of $\bar N$. To get the
full mode, which is an element of the kernel of $N$, it would be necessary to
add zeroes for the backward reactions. It is easily checked that the vectors 
defined by (\ref{sg1})-(\ref{sg4}) are solutions of the equations
(\ref{fluxb1})-(\ref{fluxb16}). It can also be shown that any solution for
which the last four components are zero is zero. Furthermore, any non-negative
solution for which at least one of the last four components in non-zero is 
a linear combination of the four solutions above and hence all the components
except the last four are zero. This verifies the defining property of 
elementary flux modes that a vector obtained by setting some, but not all, of 
the components of the generator to zero is not a solution. These vectors do 
not belong to the kernel of the incidence matrix and hence are stoichiometric 
generators. Finally, all solutions are linear combinations with non-negative 
coefficients of these generators. This follows from the facts that the 
generators are linearly independent and that the dimension of the kernel of 
$\bar N$ is four. Each of the generators is obtained by shutting off all but 
one of the output reactions.

Each stoichiometric generator defines a subnetwork obtained by setting those
reaction rates to zero for which the corresponding component of the generator
is zero. If mass action kinetics are being considered the desired reaction
rates can be set to zero by setting the corresponding reaction constants to
zero. For other kinetics some more thought is necessary. In most of the 
slow reactions we can set $v_i$ to zero by setting the corresponding 
coefficient $V_i$ to zero. The exceptions are the transport reactions to the 
cytosol. There we have three reaction rates but only one coefficient 
$V_{\rm ex}$. Here the desired reaction rates, $v_X$ can be set to zero by 
setting $K^{-1}_X$ formally to zero. In other words, in the limit certain 
constants $K_X$ tend to infinity. It also turns out to be helpful to set the
quantity $K_{i61}^{-1}$ to zero in the limit. If we make the same assumptions on 
the kinetics as in the Pettersson model we get a system of DAE corresponding 
to the subnetwork. Call the system of this type obtained from the first of the 
four generators listed above Model 4.1.1. Concretely, it is obtained from the 
Pettersson model by setting the reaction constants $k_{23}$, $k_{24}$, 
$k_{25}$, $k_{26}$, $k_{26}$, $k_{28}$ and $k_{29}$ to zero and discarding the 
variables $x_{\rm G1P}$, $x_{\rm G6P}$. A limit is considered where these reaction 
constants are multiplied by a small constant $\zeta$ and the constants 
$K_{\rm PGA}$ and $K_{\rm GAP}$ and $K_{i61}$are multiplied by $\zeta^{-1}$. 
Biologically this corresponds to a situation where PGA and GAP not only fail
to be exported but even fail to bind to the transporter and thus do 
not compete with DHAP. In a similar way it is possible to obtain an analogue of 
the hybrid model for the subnetwork. Call it Model 4.2.1. We can freeze the
concentrations of ATP and $\rm P_i$ in Models 4.1.1 and 4.2.1 to get Models
4.1.2 and 4.2.2. Model 4.1.2 is close to a modelling approach used in 
\cite{pettersson87} although in that paper no complete system of equations was 
written out. The steady states of Model 4.1.2 can be studied by following 
calculations in \cite{pettersson87}. There are two differences between 
Model 4.1.2 and the situation in \cite{pettersson87}. One of these corresponds 
to setting the coefficients $K_{i62}^{-1}$ to zero in the expression 
for $v_6$ in \cite{pettersson88} while the other has to do with the fact that
$v_{\rm PGA}$ and $v_{\rm GAP}$ are absent from Model 4.1.2.
 
The equations (\ref{Fi01})-(\ref{Fi06})
for the reaction rates in the Pettersson model are modified in
the subnetwork by the removal of $v_{\rm st}$ and $v_{\rm PGA}$. For $\zeta=0$
we have $\Phi_2=v_{\rm st}$. Following the computations done above we find that
for $\zeta=0$
\begin{equation}\label{phi1zero}
\Phi_1=\frac{Ax_{\rm DHAP}}{B+Cx_{\rm DHAP}}
-\frac{Dx_{\rm DHAP}^2}{E+Fx_{\rm DHAP}^2}
\end{equation}
where $A=V_{\rm ex}$, 
$B=K_{\rm DHAP}\left(2+\frac{x_{\rm P_{ext}}}{K_{\rm P_{ext}}}\right)
\frac{x_{\rm P_i}}{K_{\rm P_i}}$,
$C=K_{\rm DHAP}+1+\frac{x_{\rm P_{ext}}}{K_{\rm P_{ext}}}$,
$D=V_6\frac{k_6k_9}{k_7k_8}$, $E=1+K_{i62}^{-1}x_{\rm P_i}$ and 
$F=\frac{k_6k_9}{k_7k_8}$.
Using the positivity of the unknown we see that the equation $\Phi_1=0$ is 
equivalent to the quadratic equation
\begin{equation}
(AF-CD)x_{\rm DHAP}^2-BDx_{\rm DHAP}+AE=0.
\end{equation}
This equation has two positive solutions precisely when 
$AF-CD>0$ and $AE<\frac{B^2D^2}{4(AF-CD)}$. Moreover in that case the 
derivative of $\Phi_1$ is non-zero at each of those points. Parameters can be 
chosen such that these inequalities are satisfied. For instance, starting 
from arbitrary positive values of the parameters $V_6$ can be reduced so as
to ensure that the first inequality is satisfied. Then $k_6$ can be increased
to arrange that the second one is satisfied. Perturbing $\zeta$ away from zero 
and applying the implicit function theorem we see that there exist two positive 
steady states of Model 3.4.1 for suitable choices of the parameters. This 
implies in turn the existence of two positive steady states for the Pettersson
model. Summing up, we get the following result

\noindent
{\bf Theorem} There are choices of the parameters for which the Pettersson 
model has at least two positive steady states.

\section{Steady states of the Poolman model}\label{steadypo}

The equations for the steady state fluxes in the Poolman model are similar to 
those in the case of the Pettersson model. If we make use of the conservation
law for the total amount of phosphate then the only difference is an 
additional summand $v_{17}$ in the evolution equation for G1P. Note that this 
reaction rate belongs to a slow reaction. The explicit expression for  
this rate is 
$v_{17}=\frac{V_{17}x_{\rm P_i}}{x_{\rm P_i}+K_{m17}
\left(1+\frac{x_{\rm G1P}}{K_{i17}}\right)}$ (cf. \cite{poolman99}, equation 
(4.4)). The equations (54)-(58) in 
\cite{pettersson88} are replaced by
\begin{eqnarray}
&&F_1=v_{13}-v_1,\\
&&F_2=v_6-v_9-v_{\rm st}+v_{17},\\
&&F_3=v_6+2v_9-v_{13}-v_{\rm st}+v_{17},\\
&&F_4=2v_1+v_{\rm st}-v_{\rm ex}-2v_9-3v_6-v_{17},\\
&&F_5=v_{16}+v_{\rm PGA}-2v_1-v_{13}-v_{\rm st}.
\end{eqnarray}  
Correspondingly equations (42)-(47) of \cite{pettersson88} are replaced by
\begin{eqnarray}
&&v_1=v,\\
&&v_6=\frac{v}{3}+v_{\rm st}-v_{17},\\
&&v_9=\frac{v}{3},\\
&&v_{13}=v,\\
&&v_{16}=3v+v_{\rm st}-v_{\rm PGA},\\
&&v=3v_{\rm ex}+6v_{\rm st}-6v_{17}.
\end{eqnarray} 
Let us freeze ATP and $\rm P_i$ in the reduced Poolman model and call the 
result Model 4.1.1. In the frozen model $v_{17}$ is a function of $x_{\rm G1P}$ 
alone.

We now pass to a limit in a similar way to what was done for the Pettersson 
model, setting the reaction rates $v_{\rm PGA}$, $v_{\rm PGA}$ and $v_{\rm st}$ to 
zero. This time we allow $v_{17}$ to remain non-zero. The calculations are 
simplified by setting $K_{i17}^{-1}$ to zero, so that the expression for $v_{17}$
reduces to a constant. Then equation (\ref{phi1zero}) is replaced by
\begin{equation}\label{phi1zero17}
\Phi_1=\frac{Ax_{\rm DHAP}}{B+Cx_{\rm DHAP}}
-\frac{Dx_{\rm DHAP}^2}{E+Fx_{\rm DHAP}^2}\label{Phi2}-G
\end{equation}
where $G=\frac{3V_{17}x_{\rm P_i}}{x_{\rm P_i}+K_{m17}}$. If we start from a choice
of parameters which gives two positive steady states in the Pettersson model
and perturb $G$ from being zero to being positive and sufficiently small then
the function $\Phi_1$ has three non-degenerate zeroes. For a sufficiently small
perturbation of this type does not destroy the positive zeroes present for 
$G=0$ and does not make them become degenerate. On the other hand is makes
the value of $\Phi_1$ at the origin negative and this leads to a new positive
zero. This zero is a deformation of the zero which lies at the origin for
$G=0$ and so it is non-degenerate for small parameter values. It follows by
arguments similar to those in the last section that for the parameter values 
just considered the reduced Poolman model has three positive steady states.

Stoichiometric generators for the Poolman system have been studied
in \cite{poolman03} and we now consider the relation of these to the 
construction just carried out. There are modes generalizing those for the 
Pettersson model by augmenting them by a zero entry for $v_{17}$. Strangely 
these do not seem to fit with Figure 2A in \cite{poolman03} where the reaction 
with flux $v_7$ is also shown as being shut off. That figure appears to 
contradict equation (9) of \cite{pettersson88} which says that in a steady state
$v_6=v_7+v_{14}$. We can now proceed as in the last section, with $v_{17}$ 
being set to zero. We use the mode
\begin{equation}
[3\ 6\ 6\ 3\ 1\ 1\ 1\ 1\ 1\ 1\ 1\ 2\ 3\ 0\ 0\ 10\ 0\ 0\ 1\ 0\ 0].
\end{equation}
That takes us to the same subnetwork as before and the analogous arguments 
show that there are parameter values for which the reduced Poolman
model has two positive steady states. As has been shown above the 
possibility of having $v_{17}$ non-zero to get results for the reduced Poolman 
model which go beyond those obtained for the Pettersson model.

To obtain these modes it is necessary to be careful about the 
difference between $N$ and $\bar N$. This time we will make a different choice
of which reactions are considered to be in the forward direction. This results
in the reaction rates $v_{14}$ and $v_{15}$ being replaced by their
negatives. This is related to the fact that in this mode material is flowing
from starch to the sugars in the cycle. Consider the mode
\begin{equation}
[3\ 6\ 6\ 4\ 0\ 0\ 1\ 1\ 1\ 1\ 1\ 2\ 3\ 1\ 1\ 9\ 0\ 0\ 1\ 0\ 1].
\end{equation}
In this case we get a subnetwork with an inflow from starch but no outflow to 
starch. 

Note finally that in this paper we have not obtained any results on 
multiple stationary solutions for the Pettersson model itself or for the 
hybrid model. In order to do this it would be necessary to obtain 
information on the transverse eigenvalues, showing that they are all 
different from zero, at least for some values of parameters for which
multiple steady states exist for the reduced models.

\section{Outlook}

In this paper we have been concerned with a variety of mathematical models
for one biological system, the Calvin cycle. The approach has been to 
understand as much as possible about the relations between the different
models and to obtain as much information as possible about the qualitative 
behaviour of the solutions of the equations defined by these models. The 
scope was restricted to results obtained by purely analytical and rigorous
methods without any appeal to numerics or reliance on heuristics. These 
results are piecemeal and should be complemented by a better conceptual 
understanding of the key mechanisms determining the behaviour. To do this 
it makes sense to look at models which are as simple as possible.

The approach of studying the simplest possible model has been pursued by
Hahn \cite{hahn91}. His three-variable model includes the important 
phenomenon of photorespiration which is not included in the models 
discussed up to now in this paper. Because this seems to the author to be
an important direction for future developments the work of \cite{hahn91}
will now be discussed briefly. The unknowns are $x_{\rm RuBP}$, $x_{\rm PGA}$
and $x_{\rm TP}$. TP stands for 'triose phophate' and compared to the 
five-variable models it is obtained by lumping together $x_{\rm DPGA}$ and 
$x_{\rm GAP}$. $x_{\rm Ru5P}$ has been considered as an intermediate species and
discarded. Let us ignore photorespiration for the moment ($k_2=0$ in the 
notation of \cite{hahn91}). We also ignore the reaction called dark 
respiration ($k_5=0$).

The reaction from RuBP to PGA is as in Model 2.1.1 and there is a reaction 
taking PGA to TP replacing that from PGA to DPGA in Model 2.1.1. There is a 
sink reaction starting at TP. There is a reaction from TP to RuBP replacing 
that from GAP to Ru5P. The evolution equations are
\begin{eqnarray}
&&\frac{dx_{\rm RuBP}}{dt}=-k_1x_{\rm RuBP}+3k_4x_{\rm TP}^5,\\
&&\frac{dx_{\rm PGA}}{dt}=2k_1x_{\rm RuBP}-k_3x_{\rm PGA},\\
&&\frac{dx_{\rm TP}}{dt}=k_3x_{\rm PGA}-5k_4x_{\rm TP}^5-k_6x_{\rm TP}.
\end{eqnarray}
It is emphasized by Hahn that a key property which is expected from a model
is that it should have a stable positive steady state. He states that $k_1$ can
reasonably be estimated from experimental data but that $k_3$, $k_4$ and $k_6$
cannot. Thus he adopts the following strategy. It is assumed that a stable 
steady state exists in the model and the concentrations of the substances 
involved which can be measured under suitable circumstances are assumed to be 
the values in the steady state. Assuming in this way certain values for the 
coordinates of the steady state the three remaning reaction constants can be
calculated. Having obtained these values we can then ask if for those 
parameters there are other positive steady states.   

The system above has exactly one positive steady state for any values of the 
parameters. It satisfies 
$x_{\rm TP}=\left(\frac{k_1x_{\rm RuBP}}{3k_4}\right)^{\frac15}$, 
$x_{\rm PGA}=\frac{2k_1x_{\rm RuBP}}{k_3}$ and 
\begin{equation}
x_{\rm RuBP}=\left[\left(\frac{k_1}{3k_6}\right)^5\frac{3k_4}{k_1}\right]^{\frac14}.
\end{equation}
Note that in contrast to Model 2.1.1 this model always has a positive steady
state for any choice of the parameters. The difference has to do with the
fact that in the Hahn model there is no sink for $x_{\rm PGA}$. It is shown in
\cite{hahn91} that this steady state is unstable.

In the case with photorespiration it is shown in \cite{hahn91} how to reduce
the problem of finding steady states to that of solving a ninth degree
equation for one of the concentrations. Numerically it was found that for 
the biologically motivated values of the parameters this model has two steady
states. Moreover one of these is stable and the other unstable. This is
similar to the results which have proved for other models discussed in the
previous sections. In the model of \cite{hahn91} this situation is only 
possible in the presence of photorespiration. However the other models suggest
that in this respect Hahn's three-variable model is not representative of 
what happens in more detailed models. It is desirable to obtain more analytical
results on the Hahn models and their relations to other more detailed models
of the Calvin cycle. Note that in \cite{hahn84} Hahn had also previously 
studied some larger models which have rise to the models of \cite{hahn91} by
a process of simplification.

\end{document}